\documentclass[aps,prl,twocolumn,showpacs,amsmath,amsmath,amssymb,amsfonts,superscriptaddress]{revtex4}
\usepackage{amssymb}
\usepackage{graphicx}
\usepackage{dcolumn}
\usepackage{bm}
\usepackage{psfrag}
\usepackage{bbold}
\usepackage{float}
\usepackage{amsmath}

\begin{document}
\author{B.Olmos}
\author{D. Yu}
\author{I. Lesanovsky}
\affiliation{School
of Physics and Astronomy, The University of Nottingham, Nottingham,
NG7 2RD, United Kingdom}

\title{Steady state properties of a driven atomic ensemble with non-local dissipation}
\date{\today}
\begin{abstract}
The steady state of a driven dense ensemble of two-level atoms is determined from the competition of coherent laser excitation and decay that acts in a correlated way on several atoms simultaneously. We show that the presence of this non-local dissipation lifts the direct link between the density of excited atoms and the photon emission rate which is typically present when atoms decay independently. The non-locality disconnects these static and dynamic observables so that a dynamical transition in one does not necessarily imply a transition in the other. Furthermore, the collective nature of the quantum jump operators governing the non-local decay results in the formation of spatial coherence in the steady state which can be measured by analyzing solely global quantities - the photon emission rate and the density of excited atoms. The experimental realization of the system with strontium atoms in a lattice is discussed.
\end{abstract}


\maketitle
Cold atoms offer a flexible and versatile toolbox for the experimental realization of many-body quantum systems \cite{Bloch08}. They allow one to tailor a wide range of coherent dynamics by tuning trapping potentials and interactions \cite{Greiner02,Stoeferle04,Kollath07,Paredes04,Jaksch05}. Since recently, an interesting direction is pursued which aims at the engineering of dissipative dynamics in cold atomic systems \cite{Diehl08,Kraus08}. The action of an adequately tailored collective or non-local dissipation - for example governed by a master equation with jump operators acting in a correlated way on several atoms - can drive ensembles into particular steady states featuring entanglement or many-body correlations such as topological order or fermionic pairing \cite{Verstraete09,Weimer10,Diehl10,Yi12,Schirmer10,Bardyn12,Cormick13}. Moreover, the competition between coherent and engineered dissipative dynamics that is inherent to these systems can induce phase transitions in their steady state \cite{Tomadin11,Honing12,Sieberer13,Carr13}.

The experimental implementation of such systems which would permit the exploration of this intriguing non-equilibrium dynamics remains a challenge. First experiments in this direction have recently been carried out with trapped ions \cite{Barreiro11,Schindler13,Lin13}. We focus here on a system where such non-local dissipation appears naturally without the need of engineering: An ensemble of identical two-level atoms coupled to the radiation field \cite{Agarwal70,Lehmberg70}. Here, the proximity of the atoms plays a decisive role. In dense samples where the average interatomic distance is smaller than the wavelength of the considered transition, the emission and reabsorption of virtual photons among the atoms induces cooperative effects such as dipole-dipole interactions or a cooperative Lamb shift of the transition frequency \cite{Keaveney12}. Radiative decay is described by a Lindblad master equation with (non-local) jump operators that act simultaneously on several atoms instead of on individual ones. Thus, a photon emission cannot be identified with the decay of an individual atom, which leads to collective phenomena such as super- or subradiance \cite{Dicke54}.

\begin{figure}
\centering
 \includegraphics[width=1\columnwidth]{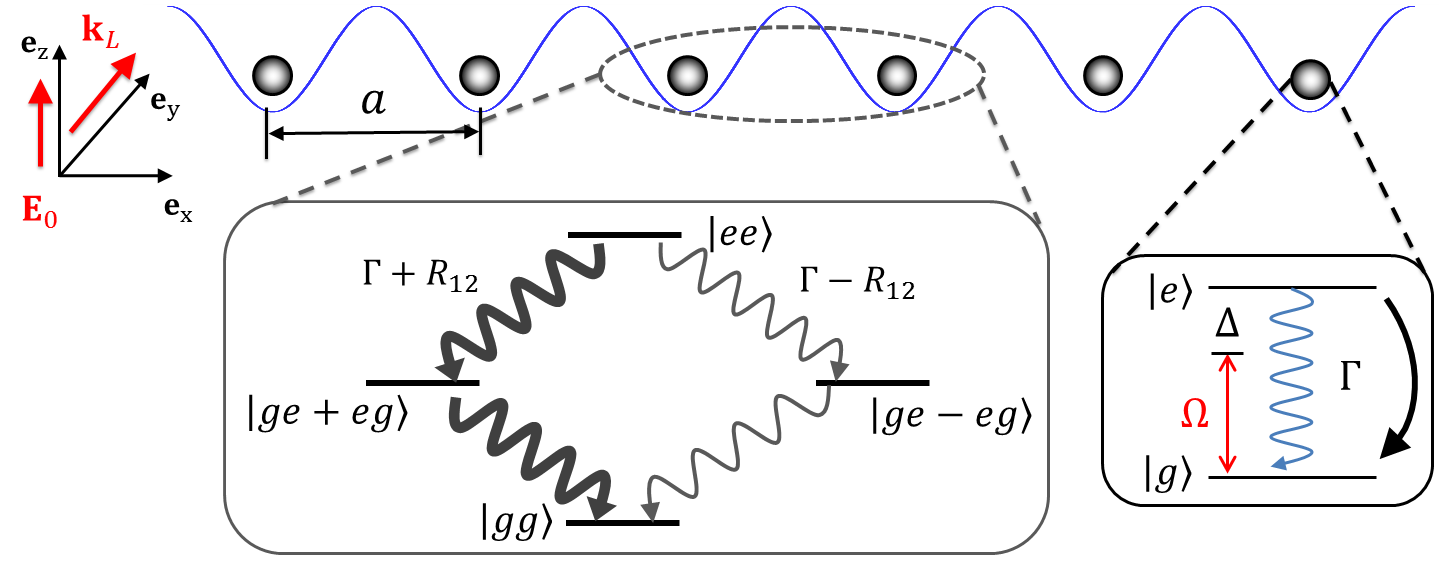}\\
 \caption{Driven ensemble of two-level atoms. The atoms are confined in a one-dimensional optical lattice with lattice constant $a$. An external laser field with amplitude $\mathbf{E}_0$ and momentum $\mathbf{k}_\mathrm{L}$ perpendicular to the lattice couples the states $\left|g\right\rangle$ and $\left|e\right\rangle$ with detuning $\Delta$ and Rabi frequency $\Omega$. The wavelength of the atomic transition $\lambda$ is much larger than $a$ and, as a consequence, interatomic interactions are induced via the exchange of virtual photons among the atoms (see text). Under these conditions the decay acquires a non-local character with the appearance of superradiant and subradiant modes, that decay at an enhanced or reduced rate ($\Gamma+R_{12}$ and $\Gamma-R_{12}$ in the two-atom case depicted).}\label{fig:scheme}
\end{figure}

In this paper, we analyze the steady state that emerges in such a dense atomic gas as a result of the competition between coherent driving and the naturally occurring non-local dissipation. Our starting point is the discussion of the behavior of the density of excited atoms in the system. We find signatures of a dynamical first order transition where the number distribution of excited atoms becomes bimodal \cite{Malossi13}. In systems with local dissipation, i.e. where each decay event can be associated with an individual atom, such transition in the \emph{static} observable of the system is often accompanied by a transition in the photon emission rate into the bath \cite{Lee11,Lee12,Ates12-2}, which can be regarded as a \emph{dynamical} order parameter \cite{Lesanovsky12b}. The non-local character of the dissipation, however, lifts this connection and changes in the statics are in general not directly visible in the photon count. Moreover, we show that in the region where the dynamical transition in the density takes place, the many-body steady state features spatial phase coherence. This phase coherence can be directly quantified from the measurement of two global quantities - the mean density of excited atoms and the average photon emission rate. Finally, we describe how all these features can be experimentally explored with strontium atoms in a lattice \cite{Olmos13}.

The setup we have in mind is depicted in Fig. \ref{fig:scheme}. Here, we are considering an ensemble of $N$ two-level atoms, each of which is initially in the electronic ground state $\left|g\right\rangle$. An external laser field is applied to couple $\left|g\right\rangle$ to the excited state $\left|e\right\rangle$. The wavelength of the corresponding transition is $\lambda$, and the radiative decay of an isolated atom from $\left|e\right\rangle$ to $\left|g\right\rangle$ takes place with decay rate $\Gamma$. All atoms are confined in a deep one-dimensional (1D) optical lattice in the Mott insulator state with one atom per site. The lattice potential is state-independent and the distance between adjacent sites $a$ is much smaller than the wavelength $\lambda$.

The density matrix $\rho$ of the atomic ensemble evolves under the Lindblad master equation
\begin{equation}\label{eqn:Master}
 \dot{\rho}=-\frac{i}{\hbar}\left[H,\rho\right]+{\cal D}(\rho).
\end{equation}
The first term on the right side describes the coherent time-evolution governed by the many-body Hamiltonian
\begin{equation}\label{eqn:Hamiltonian}
H=\hbar\sum_\alpha\!\left[-\Delta b^\dag_{\alpha}b_\alpha+\Omega\left(b_\alpha^\dag + b_\alpha\right)\right]+\hbar\sum_{\alpha\neq\beta}V_{\alpha\beta}b_\alpha^\dag b_\beta,
\end{equation}
with $b_\alpha\equiv\left|g\right>_\alpha\!\left<e\right|$. The detuning between the frequency of the laser field $\omega_\mathrm{L}$ and the atomic transition $\omega_\mathrm{a}$ is denoted by $\Delta=\omega_\mathrm{L}-\omega_\mathrm{a}$. The Rabi frequency is given by $\Omega=pE_{0}/2\hbar$, with $E_0$ being the amplitude of the laser field and $p$ the atomic transition dipole moment. The interatomic interactions that result from the exchange of virtual photons are characterized by the matrix elements $V_{\alpha\beta}=3\Gamma/4(-\cos\kappa_{\alpha\beta}/\kappa_{\alpha\beta} +\sin\kappa_{\alpha\beta}/\kappa_{\alpha\beta}^{2} +\cos\kappa_{\alpha\beta}/\kappa_{\alpha\beta}^{3})$, where $\kappa_{\alpha\beta}\equiv 2\pi r_{\alpha\beta}/\lambda$ with $r_{\alpha\beta}$ being the separation between the $\alpha$-th and $\beta$-th atoms \cite{Lehmberg70}. In traditional lattice setups, the wavelength of the transition and the lattice constant are of the same order, i.e. $a/\lambda\gtrsim 1$. Here, the value of $V_{\alpha\beta}$ is in general close to zero, i.e. the coherent dipole-dipole interaction induced by the photon emission is negligible. However, in the regime we consider here ($a/\lambda\ll1$) the interaction between atoms separated by a few sites can be approximated by a $1/r^3$ potential, with $r$ being the distance between the atoms (for details on the experimental implementation see further below).

The second term in the right hand side of eq. (\ref{eqn:Master}) is the dissipator, which describes incoherent transitions in the atomic ensemble caused by the coupling to the radiation field:
\begin{equation}\label{eqn:Liouvillian}
    {\cal D}(\rho)=\sum_{\alpha,\beta}R_{\alpha\beta}\left[b_\alpha\rho b_\beta^\dag- \frac{1}{2}\left\{b_\alpha^\dag b_\beta,\rho\right\}\right],
\end{equation}
with the matrix elements $R_{\alpha\beta}=3\Gamma/2 (\sin\kappa_{\alpha\beta}/\kappa_{\alpha\beta} +\cos\kappa_{\alpha\beta}/\kappa_{\alpha\beta}^{2} -\sin\kappa_{\alpha\beta}/\kappa_{\alpha\beta}^{3})$. The non-local character of the dissipator becomes apparent when it is brought into Lindblad form by introducing the collective jump operators $J_m=\sum_\alpha X_{\alpha m}b_\alpha$, given by superpositions of the local (lowering) operators $b_\alpha$. The matrix $X$ contains the eigenvectors of $R$. The dissipator then becomes ${\cal D}(\rho)=\sum_{m}\gamma_m[J_m\rho J_m^\dag- (1/2)\left\{J_m^\dag J_m,\rho\right\}]$. For $a/\lambda \gtrsim 1$ the matrix $R$ is nearly diagonal so that $\gamma_m\to\Gamma$ and each collective jump operator $J_m$ corresponds to a single local lowering operator $b_m$ with $m=1\dots N$. They become non-local when $a/\lambda\ll 1$ and $R$ accumulates weight on the super- and sub-diagonals. In this regime each of the decay processes associated with the jump operators $J_m$ has in general a different decay rate $\gamma_m$ that can be vastly distinct from $\Gamma$ \cite{Lehmberg70,Agarwal70}.

\begin{figure}
\centering
  \includegraphics[width=\columnwidth]{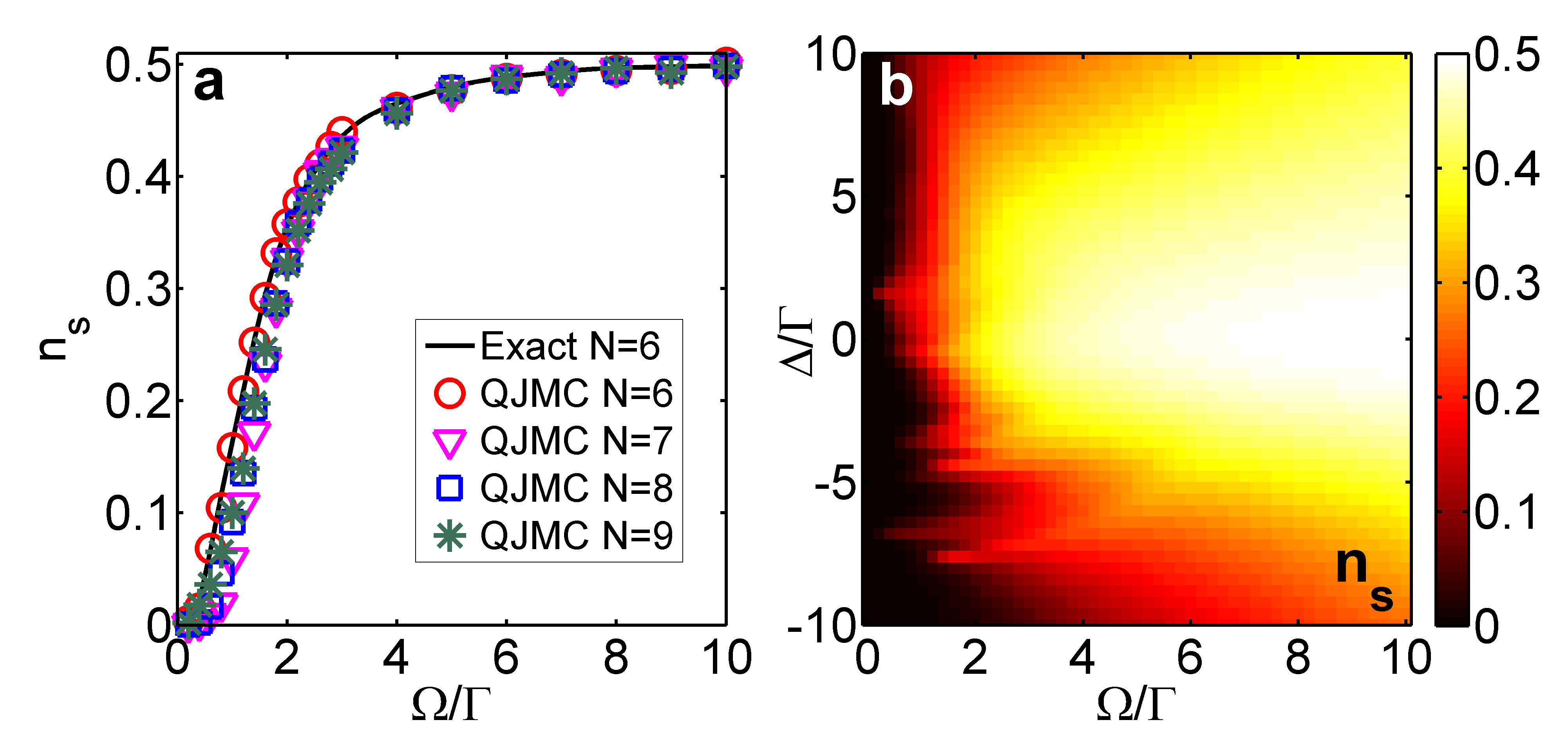}\\
  \caption{\textbf{a}: Excitation density $n_\mathrm{s}$ for different $N$ obtained from the QJMC simulations with $\Delta=0$ and $a/\lambda=0.08$. For comparison, we also show the corresponding numerically exact solutions for $N=6$ given by the solid line. \textbf{b}: Numerically exact stationary solution of the excitation density $n_\mathrm{s}$ as a function of $\Omega/\Gamma$ and $\Delta/\Gamma$ for a system with $N=6$ atoms.}\label{fig:Exact_MC}
\end{figure}

Let us now investigate the steady state $\rho_\mathrm{s}$ of Eq. (\ref{eqn:Master}) that emerges as a result of the coherent driving and this non-local dissipation. We do this by means of two approaches. The first one is to calculate $\rho_\mathrm{s}$ by numerically solving $-(i/\hbar)\left[H,\rho_\mathrm{s}\right]+\mathcal{D}(\rho_\mathrm{s})=0$. The second one is to apply a Quantum Jump Monte Carlo (QJMC) method~\cite{Molmer93,Dalibard92} in order to obtain a representative sampling of $\rho_\mathrm{s}$. The QMJC method has the advantage that it can in general be applied to larger systems and that it generates (quantum jump) trajectories which are directly comparable to experimental records. In our simulations, we generate an ensemble of $2\times10^{3}$ trajectories of length $50\Gamma^{-1}$. In Fig. \ref{fig:Exact_MC}a we show the density of excited atoms $n_\mathrm{s}=\sum_\alpha \mathrm{Tr} (b^\dagger_\alpha b_\alpha\,\rho_\mathrm{s})/N$ as a function of $\Omega/\Gamma$ ($\Delta=0$) for various system sizes. We observe a good agreement between the results using the two methods for $N=6$ atoms and moreover find that the qualitative behavior of $n_\mathrm{s}$ does not change for the system sizes shown: There is a sharp crossover at $\Omega\approx\Gamma$ from $n_\mathrm{s}\approx 0$ at $\Omega\ll\Gamma$ to $n_\mathrm{s}\approx 1/2$ for $\Omega\gg\Gamma$.

Before continuing let us briefly digress to discuss a peculiarity of the presence of non-local dissipation which is the emergence of subradiant states \cite{Dicke54}. These collective atomic states are many-body states which weakly couple to the dissipation and the coherent interaction and whose decay rates $\gamma_m$ are several orders of magnitude smaller than the single atom one, $\Gamma$. Thus, in principle it could take a time much greater than $[\min(\gamma_m)]^{-1}$ to confidently reach the stationary state with the QJMC method (and in an experiment). However, the initial state considered (all atoms in the ground state) has a negligible overlap with these subradiant states and in the course of the evolution their population is negligible. Thus, indeed the state approached after $50\Gamma^{-1}$ in the QJMC simulations (Fig. \ref{fig:Exact_MC}a) can, for all practical purposes, be considered as the system's steady state.

Let us now explore the excitation density as a function of $\Delta/\Gamma$ and $\Omega/\Gamma$. The numerically exact solution for $N=6$ depicted in Figure~\ref{fig:Exact_MC}b already provides a first insight into what is expected for larger systems: The phase diagram can be divided into two regions - one with low and one with high excitation density. When $\Omega\ll\Gamma$, the dissipation dominates the dynamics and thus in the stationary state very few atoms are excited to the $\left|e\right\rangle$ state, hence $n_\mathrm{s}\approx0$. The analysis in the case of $\Omega\gg\Gamma$ is more involved. In order to describe the system in this limit, we go to a rotating frame by means of a unitary transformation $U=\prod_\alpha\exp{-i\Omega t\left(b_\alpha^\dag+b_\alpha\right)}$ and neglect the fast rotating terms with frequency $\Omega$. Following this procedure one can show that the rotated dissipator $U^\dag {\cal D}(\rho)U$ is a sum of three terms. Each of these terms has the same formal expression as (\ref{eqn:Liouvillian}) but substituting the operator $b_\alpha$ by $1/2(b_\alpha+b_\alpha^\dag)$, $i/(2\sqrt{2})(b_\alpha-b_\alpha^\dag)$ and $1/(2\sqrt{2})(2b_\alpha^\dag b_\alpha-1)$. As one can see, these are all hermitian operators, and hence one can prove that the stationary solution of the system in the limit of $\Omega\gg\Gamma$ is the completely mixed state \cite{Kraus08}. This is indeed  consistent with the value of the excitation density $n_\mathrm{s}\approx1/2$ found here.

\begin{figure}
\centering
  \includegraphics[width=\columnwidth]{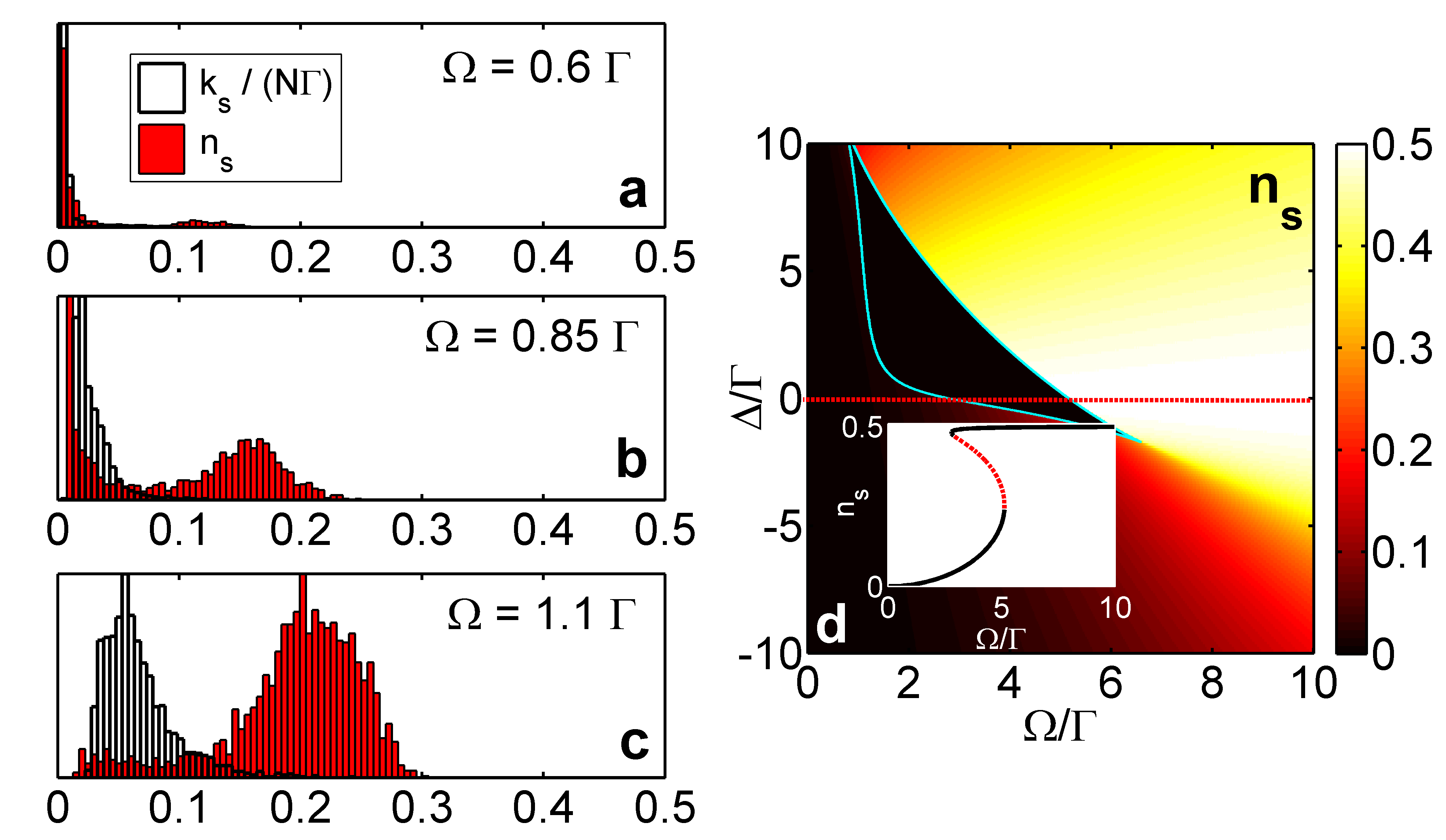}\\
  \caption{\textbf{a-c}: Distribution functions of the excitation density and emission rate obtained from the QJMC simulations for three different values of $\Omega/\Gamma$ across the crossover region in the resonant case (cf. Fig. \ref{fig:Exact_MC}a) for a system with $N=12$ atoms and $a/\lambda=0.08$. The histograms display a bimodal behavior of the excitation density while the emission rate is unimodal. \textbf{d}: The stationary value of the excitation density $n_\mathrm{s}$ as a function of $\Delta/\Gamma$ and $\Omega/\Gamma$ derived from the mean-field theory. The curve delimited by a solid line in the $\Delta-\Omega$ plane represents the boundary of the bistable region. In the inset, the black solid lines correspond to the stable solutions while the red dashed lines denote the unstable ones in the case $\Delta=0$.}\label{fig:hists}
\end{figure}

Next, we elaborate on the consequences arising from the non-local character of the dissipation, which is a very distinctive property of our system that separates our study from related previous ones, e.g. Refs. \cite{Lee11,Lee12,Ates12-2}. In a general system described by a Lindblad master equation with a set of jump operators $J_m$ and rates $\gamma_m$, the average emission rate $k_\mathrm{s}$ into the bath (summed over all possible decay channels) in the stationary state is given by the expectation value
\begin{equation}\label{eqn:ks}
    k_\mathrm{s}=\sum_m \gamma_m \langle J_m^\dag J_m\rangle_\mathrm{s},
\end{equation}
where $\left<...\right>_s\equiv\mathrm{Tr}(...\rho_\mathrm{s})$. As we pointed out before, in the case of an ensemble of $N$ two-level atoms in which each atom couples independently to a zero temperature bath with rate $\Gamma$ the jump operators act on individual atoms, i.e. $J_m\to b_m$ with $m=1\dots N$. Here, the emission rate is proportional to the mean excitation density, i.e. $k_\mathrm{s}=N \Gamma n_\mathrm{s}$. Hence, changes in the static observable $n_\mathrm{s}$, for instance phase transitions, become manifest in a changing mean rate of emitted photons, as discussed for example in Refs. \cite{Lee11,Lee12,Ates12-2}. However, in general there is no such simple connection between these static and dynamic observables \cite{Lesanovsky12b}. In particular, the proportionality relation $k_\mathrm{s}=N \Gamma n_\mathrm{s}$ between the mean values does not hold in the dissipative many-body system we are considering here, as the jump operators are non-local. This becomes even clearer at the level of the full distribution functions of the corresponding observables: In Figs.~\ref{fig:hists}a, b, and c, we show the distributions of the excitation density and photon emission rate for three different values of $\Omega/\Gamma$ in the crossover region ($\Omega\sim\Gamma$) with $\Delta=0$. These data have been obtained from QJMC simulations of a system with $N=12$ atoms.

We first focus in the distribution of excitation density: For low $\Omega/\Gamma$ (Fig. \ref{fig:hists}a) the distribution is unimodal with the maximum being near zero. As one increases the Rabi frequency, the distribution becomes bimodal (Fig. \ref{fig:hists}b). Finally, Fig. \ref{fig:hists}c shows that the distribution for larger $\Omega$ becomes unimodal again with the maximum being at a density near $0.2$. These results suggest a dynamical first order transition in the density of excited atoms. This conclusion is corroborated by a mean field treatment of the master equation (\ref{eqn:Master}): Here we find that the steady state value $n_\mathrm{s}$ is determined by a cubic equation. In Fig. \ref{fig:hists}d we show the resulting mean field phase diagram. Note that the appearance is similar to the exact solution for small systems shown in Fig. \ref{fig:Exact_MC}b. For most parameter regimes, the mean field equation possesses a unique solution: For $\Omega\ll\Gamma$, the excitation probability is low and, in contrast, when $\Omega\gg\Gamma$ the excitation probability approaches $1/2$. In particular, in the latter regime we can approximate $n_\mathrm{s}\approx 1/2-(\Gamma^2+4\Delta^2)/(8\Omega^2)$. However, there is one region in parameter space - delimited by the solid lines - in which the mean field equation has two stable solutions which can be interpreted as two coexisting steady states that correspond to a low and a high excitation density as it can be seen in the inset in Fig. \ref{fig:hists}d. This coexistence becomes directly manifest in the QJMC simulations through the bimodality of the histogram (Fig. \ref{fig:hists}b). In other works where the dissipation was localized to individual atoms, this bimodality translated into a strongly intermittent photon emission \cite{Lee11,Lee12,Ates12-2}, i.e., also the distribution of the photon emission rate was bimodal. Figs. \ref{fig:hists}a b and c show that this connection is not present here as in fact the photon emission rate has always a unimodal character, clearly distinct from the distribution of the excitation density.

\begin{figure}
\centering
  \includegraphics[width=\columnwidth]{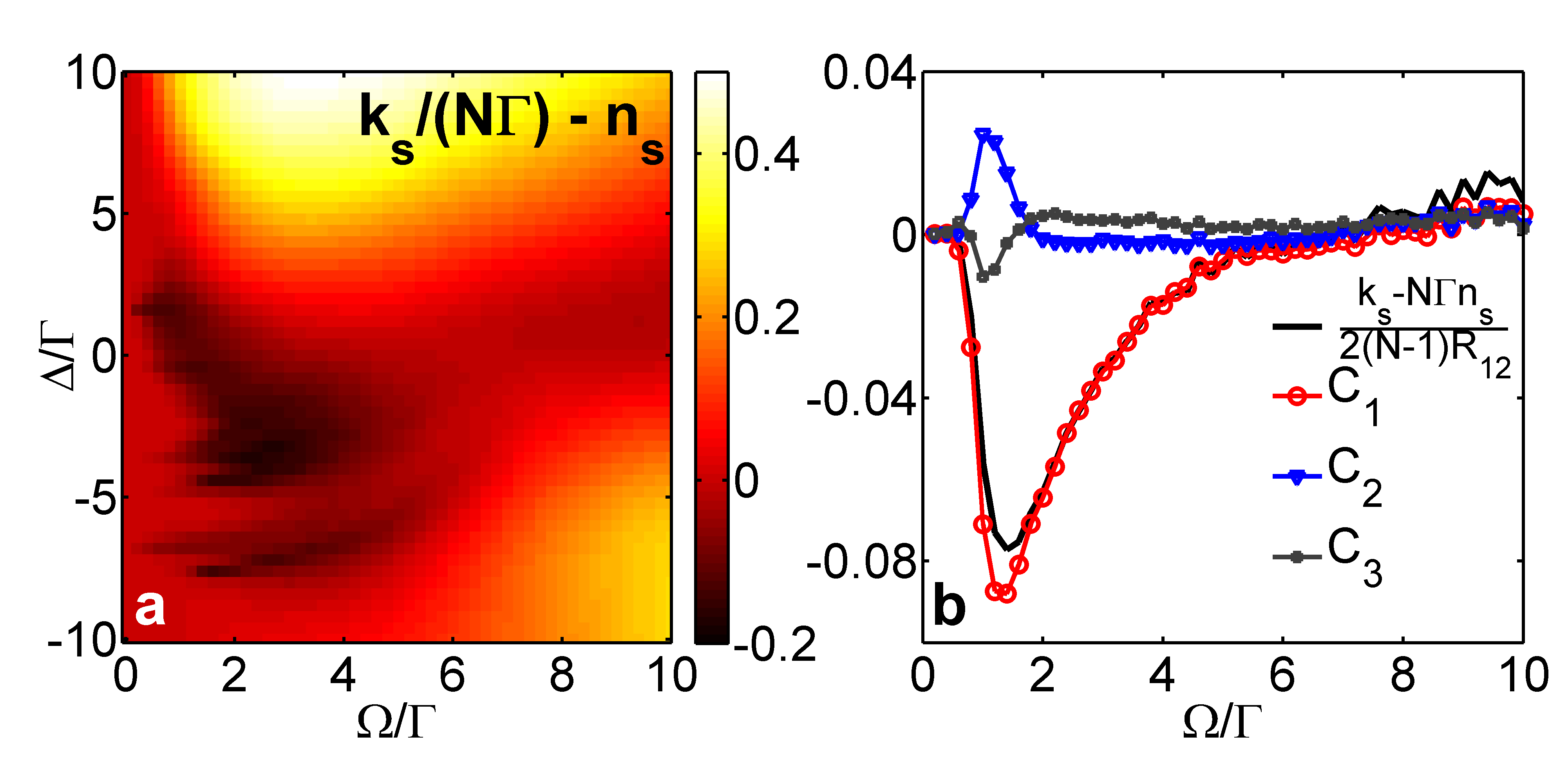}\\
  \caption{\textbf{a}: Numerically exact value of $k_\mathrm{s}/N\Gamma-n_\mathrm{s}$ for a system with $N=6$ atoms as a function of $\Delta/\Gamma$ and $\Omega/\Gamma$ with $a/\lambda=0.08$. \textbf{b}: The average interatomic coherences $C_{d}$ ($d=1,2,3$) derived from the QJMC simulation for a system with $N=8$ atoms as a function of $\Omega/\Gamma$ in the resonant case.}\label{fig:correlations}
\end{figure}
The absence of such trivial connection has actually an interesting application: It can be used to extract information on the spatial coherence in the stationary state by means of global measurements. This is established through Eq. (\ref{eqn:ks}), which connects the photon emission rate $k_\mathrm{s}$ to the excitation probability $n_\mathrm{s}$ and the spatial coherences in the stationary state $\langle b_\alpha^\dag b_\beta\rangle_\mathrm{s}$:
\begin{equation}\label{eqn:actandcorr}
    k_\mathrm{s}-N\Gamma n_\mathrm{s}=\!\sum_{\alpha\neq \beta}\!R_{\alpha\beta}\langle b_\alpha^\dag b_\beta\rangle_\mathrm{s}=2\!\sum_{d}\!R_{11+d}(N-d)C_d.
\end{equation}
Here we have introduced the average of the real part of the spatial coherence between atoms separated by $d$ sites $C_d=\sum_\alpha\Re{\langle b_\alpha^\dag b_{\alpha+d}\rangle_\mathrm{s}}/(N-d)$. Figure \ref{fig:correlations}a displays the value of $k_\mathrm{s}/N\Gamma-n_\mathrm{s}$ as a function of $\Omega/\Gamma$ and $\Delta/\Gamma$ for the numerically exact solution for $N=6$ atoms. One can observe that the difference is zero in the two limiting cases $\Omega\ll\Gamma$ and $\Omega\gg\Gamma$. However, in the region where $\Omega\approx\Gamma$, i.e. where we also observe the bimodal behavior of the excitation density, the difference is in general non-zero. Thus, from Eq. (\ref{eqn:actandcorr}) we can infer that in this regime the competition between coherent driving and the non-local dissipation leads to a steady state which features phase coherence between spatially separated atoms.

Let us focus on the case of resonant laser excitation, $\Delta=0$. In Figure~ \ref{fig:correlations}b we show the coherence between atoms at different distances $C_d$ ($d=1,2,3$) corresponding to a system with $N=8$ atoms. The nearest neighbor coherence $C_1$ acquires a negative value within the crossover region $\Omega\approx\Gamma$. Moreover, the value of the next nearest neighbor coherence $C_2$ is positive and smaller than $C_1$. The coherence clearly decays with the distance between the atoms, dying out approximately after next nearest neighbors, with $C_3$ having a small negative value. For the parameter regime used here the nearest neighbors coherence $C_1$ is thus the largest contribution to the sum in Eq. (\ref{eqn:actandcorr}). We can therefore approximate the nearest-neighbor phase coherence by $C_1\approx \frac{k_\mathrm{s}-N\Gamma n_\mathrm{s}}{2\left(N-1\right)R_{12}}$. Hence, in our system the measurement of the difference between the excitation probability and emission rate maps directly into the nearest neighbor spatial coherence of the many-body steady state.

Let us finally discuss the experimental realization of the proposed setup. The main difficulty to overcome is to achieve a situation in which the ratio between lattice spacing and photon wavelength $a/\lambda$ is much smaller than one. However, these conditions can be reached in a system of cold bosonic strontium atoms proposed in \cite{Olmos13}. Here, the ground and excited states of the model two-level atom are represented by the metastable $\left|(5s5p)^{3}\!P_{0}\right\rangle$ and $\left|(5s4d)^{3}\!D_1(m=0)\right\rangle$ triplet states, respectively. The wavelength and dipole moment of this transition are $\lambda=2.6$ $\mu$m and $p=4.03$ Debye, respectively. Both internal states can be trapped simultaneously by an optical lattice at a magic wavelength $\lambda_{b}=412.8$ nm \cite{Olmos13}, such that the lattice constant is $a=206.4$ nm. Thus, the ratio between lattice constant and wavelength is here $a/\lambda=0.08$. This value has been used for all numerical simulations in this paper. Since $\left|^{3}\!D_1\right\rangle$ decays preferentially to $\left|^{3}\!P_0\right\rangle$ with branching ratio around 60\% \cite{Zhou10} and spontaneous emission rate $\Gamma=290\times10^3$ $s^{-1}$, this system can indeed be regarded as a close approximation to an open one-dimensional many-body system composed of laser-driven two-level atoms.

In summary, we have studied the steady state of a driven ensemble of two-level atoms subject to naturally arising non-local dissipation. In this system local static and dynamical observables are not directly connected as in the simpler case where atoms are coupled to individual localized baths. This leads to a steady state which, in certain parameter regimes, exhibits spatial phase coherence between atoms. The non-equilibrium physics discussed here can be probed in lattices of Sr atoms. It will be interesting in the future to analyze the usefulness of the emerging entangled states for practical applications such as quantum information processing and precision measurements \cite{Krischek11,Ostermann13}.

This research has been supported by the EU-FET grant QuILMI 295293. B.O. acknowledges funding from University of Nottingham. Useful discussions with C. Ates, M. Hush and S. Genway are gratefully acknowledged.


\begin{thebibliography}{37}
\expandafter\ifx\csname natexlab\endcsname\relax\def\natexlab#1{#1}\fi
\expandafter\ifx\csname bibnamefont\endcsname\relax
  \def\bibnamefont#1{#1}\fi
\expandafter\ifx\csname bibfnamefont\endcsname\relax
  \def\bibfnamefont#1{#1}\fi
\expandafter\ifx\csname citenamefont\endcsname\relax
  \def\citenamefont#1{#1}\fi
\expandafter\ifx\csname url\endcsname\relax
  \def\url#1{\texttt{#1}}\fi
\expandafter\ifx\csname urlprefix\endcsname\relax\def\urlprefix{URL }\fi
\providecommand{\bibinfo}[2]{#2}
\providecommand{\eprint}[2][]{\url{#2}}

\bibitem[{\citenamefont{Bloch et~al.}(2008)\citenamefont{Bloch, Dalibard, and
  Zwerger}}]{Bloch08}
\bibinfo{author}{\bibfnamefont{I.}~\bibnamefont{Bloch}},
  \bibinfo{author}{\bibfnamefont{J.}~\bibnamefont{Dalibard}}, \bibnamefont{and}
  \bibinfo{author}{\bibfnamefont{W.}~\bibnamefont{Zwerger}},
  \bibinfo{journal}{Rev. Mod. Phys.} \textbf{\bibinfo{volume}{80}},
  \bibinfo{eid}{885} (\bibinfo{year}{2008}).

\bibitem[{\citenamefont{Greiner et~al.}(2002)\citenamefont{Greiner, Mandel,
  Esslinger, H\"ansch, and Bloch}}]{Greiner02}
\bibinfo{author}{\bibfnamefont{M.}~\bibnamefont{Greiner}},
  \bibinfo{author}{\bibfnamefont{O.}~\bibnamefont{Mandel}},
  \bibinfo{author}{\bibfnamefont{T.}~\bibnamefont{Esslinger}},
  \bibinfo{author}{\bibfnamefont{T.~W.} \bibnamefont{H\"ansch}},
  \bibnamefont{and} \bibinfo{author}{\bibfnamefont{I.}~\bibnamefont{Bloch}},
  \bibinfo{journal}{Nature} \textbf{\bibinfo{volume}{415}}, \bibinfo{pages}{39}
  (\bibinfo{year}{2002}).

\bibitem[{\citenamefont{St\"oferle et~al.}(2004)\citenamefont{St\"oferle,
  Moritz, Schori, K\"ohl, and Esslinger}}]{Stoeferle04}
\bibinfo{author}{\bibfnamefont{T.}~\bibnamefont{St\"oferle}},
  \bibinfo{author}{\bibfnamefont{H.}~\bibnamefont{Moritz}},
  \bibinfo{author}{\bibfnamefont{C.}~\bibnamefont{Schori}},
  \bibinfo{author}{\bibfnamefont{M.}~\bibnamefont{K\"ohl}}, \bibnamefont{and}
  \bibinfo{author}{\bibfnamefont{T.}~\bibnamefont{Esslinger}},
  \bibinfo{journal}{Phys. Rev. Lett.} \textbf{\bibinfo{volume}{92}},
  \bibinfo{pages}{130403} (\bibinfo{year}{2004}).

\bibitem[{\citenamefont{Kollath et~al.}(2007)\citenamefont{Kollath,
  L\"{a}uchli, and Altman}}]{Kollath07}
\bibinfo{author}{\bibfnamefont{C.}~\bibnamefont{Kollath}},
  \bibinfo{author}{\bibfnamefont{A.~M.} \bibnamefont{L\"{a}uchli}},
  \bibnamefont{and} \bibinfo{author}{\bibfnamefont{E.}~\bibnamefont{Altman}},
  \bibinfo{journal}{Phys. Rev. Lett.} \textbf{\bibinfo{volume}{98}},
  \bibinfo{eid}{180601} (\bibinfo{year}{2007}).

\bibitem[{\citenamefont{Paredes et~al.}(2004)\citenamefont{Paredes, Widera,
  Murg, Mandel, F\"olling, Cirac, Shlyapnikov, H\"ansch, and
  Bloch}}]{Paredes04}
\bibinfo{author}{\bibfnamefont{B.}~\bibnamefont{Paredes}},
  \bibinfo{author}{\bibfnamefont{A.}~\bibnamefont{Widera}},
  \bibinfo{author}{\bibfnamefont{V.}~\bibnamefont{Murg}},
  \bibinfo{author}{\bibfnamefont{O.}~\bibnamefont{Mandel}},
  \bibinfo{author}{\bibfnamefont{S.}~\bibnamefont{F\"olling}},
  \bibinfo{author}{\bibfnamefont{J.}~\bibnamefont{Cirac}},
  \bibinfo{author}{\bibfnamefont{G.}~\bibnamefont{Shlyapnikov}},
  \bibinfo{author}{\bibfnamefont{T.}~\bibnamefont{H\"ansch}}, \bibnamefont{and}
  \bibinfo{author}{\bibfnamefont{I.}~\bibnamefont{Bloch}},
  \bibinfo{journal}{Nature} \textbf{\bibinfo{volume}{429}},
  \bibinfo{pages}{277} (\bibinfo{year}{2004}).

\bibitem[{\citenamefont{Jaksch and Zoller}(2005)}]{Jaksch05}
\bibinfo{author}{\bibfnamefont{D.}~\bibnamefont{Jaksch}} \bibnamefont{and}
  \bibinfo{author}{\bibfnamefont{P.}~\bibnamefont{Zoller}},
  \bibinfo{journal}{Annals of Physics} \textbf{\bibinfo{volume}{315}},
  \bibinfo{pages}{52 } (\bibinfo{year}{2005}).

\bibitem[{\citenamefont{Diehl et~al.}(2008)\citenamefont{Diehl, Micheli,
  Kantian, Kraus, B\"uchler, and Zoller}}]{Diehl08}
\bibinfo{author}{\bibfnamefont{S.}~\bibnamefont{Diehl}},
  \bibinfo{author}{\bibfnamefont{A.}~\bibnamefont{Micheli}},
  \bibinfo{author}{\bibfnamefont{A.}~\bibnamefont{Kantian}},
  \bibinfo{author}{\bibfnamefont{B.}~\bibnamefont{Kraus}},
  \bibinfo{author}{\bibfnamefont{H.}~\bibnamefont{B\"uchler}},
  \bibnamefont{and} \bibinfo{author}{\bibfnamefont{P.}~\bibnamefont{Zoller}},
  \bibinfo{journal}{Nat. Phys} \textbf{\bibinfo{volume}{4}},
  \bibinfo{pages}{878} (\bibinfo{year}{2008}).

\bibitem[{\citenamefont{Kraus et~al.}(2008)\citenamefont{Kraus, B\"uchler,
  Diehl, Kantian, Micheli, and Zoller}}]{Kraus08}
\bibinfo{author}{\bibfnamefont{B.}~\bibnamefont{Kraus}},
  \bibinfo{author}{\bibfnamefont{H.~P.} \bibnamefont{B\"uchler}},
  \bibinfo{author}{\bibfnamefont{S.}~\bibnamefont{Diehl}},
  \bibinfo{author}{\bibfnamefont{A.}~\bibnamefont{Kantian}},
  \bibinfo{author}{\bibfnamefont{A.}~\bibnamefont{Micheli}}, \bibnamefont{and}
  \bibinfo{author}{\bibfnamefont{P.}~\bibnamefont{Zoller}},
  \bibinfo{journal}{Phys. Rev. A} \textbf{\bibinfo{volume}{78}},
  \bibinfo{pages}{042307} (\bibinfo{year}{2008}).

\bibitem[{\citenamefont{Verstraete et~al.}(2009)\citenamefont{Verstraete, Wolf,
  and Cirac}}]{Verstraete09}
\bibinfo{author}{\bibfnamefont{F.}~\bibnamefont{Verstraete}},
  \bibinfo{author}{\bibfnamefont{M.}~\bibnamefont{Wolf}}, \bibnamefont{and}
  \bibinfo{author}{\bibfnamefont{J.}~\bibnamefont{Cirac}},
  \bibinfo{journal}{Nat. Phys} \textbf{\bibinfo{volume}{5}},
  \bibinfo{pages}{633} (\bibinfo{year}{2009}).

\bibitem[{\citenamefont{Weimer et~al.}(2010)\citenamefont{Weimer, M\"{u}ller,
  Lesanovsky, Zoller, and B\"{u}chler}}]{Weimer10}
\bibinfo{author}{\bibfnamefont{H.}~\bibnamefont{Weimer}},
  \bibinfo{author}{\bibfnamefont{M.}~\bibnamefont{M\"{u}ller}},
  \bibinfo{author}{\bibfnamefont{I.}~\bibnamefont{Lesanovsky}},
  \bibinfo{author}{\bibfnamefont{P.}~\bibnamefont{Zoller}}, \bibnamefont{and}
  \bibinfo{author}{\bibfnamefont{H.~P.} \bibnamefont{B\"{u}chler}},
  \bibinfo{journal}{Nature Physics} \textbf{\bibinfo{volume}{6}},
  \bibinfo{pages}{382} (\bibinfo{year}{2010}).

\bibitem[{\citenamefont{Diehl et~al.}(2010)\citenamefont{Diehl, Yi, Daley, and
  Zoller}}]{Diehl10}
\bibinfo{author}{\bibfnamefont{S.}~\bibnamefont{Diehl}},
  \bibinfo{author}{\bibfnamefont{W.}~\bibnamefont{Yi}},
  \bibinfo{author}{\bibfnamefont{A.~J.} \bibnamefont{Daley}}, \bibnamefont{and}
  \bibinfo{author}{\bibfnamefont{P.}~\bibnamefont{Zoller}},
  \bibinfo{journal}{Phys. Rev. Lett.} \textbf{\bibinfo{volume}{105}},
  \bibinfo{pages}{227001} (\bibinfo{year}{2010}).

\bibitem[{\citenamefont{Yi et~al.}(2012)\citenamefont{Yi, Diehl, Daley, and
  Zoller}}]{Yi12}
\bibinfo{author}{\bibfnamefont{W.}~\bibnamefont{Yi}},
  \bibinfo{author}{\bibfnamefont{S.}~\bibnamefont{Diehl}},
  \bibinfo{author}{\bibfnamefont{A.~J.} \bibnamefont{Daley}}, \bibnamefont{and}
  \bibinfo{author}{\bibfnamefont{P.}~\bibnamefont{Zoller}},
  \bibinfo{journal}{New Journal of Physics} \textbf{\bibinfo{volume}{14}},
  \bibinfo{pages}{055002} (\bibinfo{year}{2012}).

\bibitem[{\citenamefont{Schirmer and Wang}(2010)}]{Schirmer10}
\bibinfo{author}{\bibfnamefont{S.~G.} \bibnamefont{Schirmer}} \bibnamefont{and}
  \bibinfo{author}{\bibfnamefont{X.}~\bibnamefont{Wang}},
  \bibinfo{journal}{Phys. Rev. A} \textbf{\bibinfo{volume}{81}},
  \bibinfo{pages}{062306} (\bibinfo{year}{2010}).

\bibitem[{\citenamefont{Bardyn et~al.}(2012)\citenamefont{Bardyn, Baranov,
  Rico, \ifmmode \dot{I}\else \.{I}\fi{}mamo\ifmmode~\breve{g}\else
  \u{g}\fi{}lu, Zoller, and Diehl}}]{Bardyn12}
\bibinfo{author}{\bibfnamefont{C.-E.} \bibnamefont{Bardyn}},
  \bibinfo{author}{\bibfnamefont{M.~A.} \bibnamefont{Baranov}},
  \bibinfo{author}{\bibfnamefont{E.}~\bibnamefont{Rico}},
  \bibinfo{author}{\bibfnamefont{A.}~\bibnamefont{\ifmmode \dot{I}\else
  \.{I}\fi{}mamo\ifmmode~\breve{g}\else \u{g}\fi{}lu}},
  \bibinfo{author}{\bibfnamefont{P.}~\bibnamefont{Zoller}}, \bibnamefont{and}
  \bibinfo{author}{\bibfnamefont{S.}~\bibnamefont{Diehl}},
  \bibinfo{journal}{Phys. Rev. Lett.} \textbf{\bibinfo{volume}{109}},
  \bibinfo{pages}{130402} (\bibinfo{year}{2012}).

\bibitem[{\citenamefont{Cormick et~al.}(2013)\citenamefont{Cormick, Bermudez,
  Huelga, and Plenio}}]{Cormick13}
\bibinfo{author}{\bibfnamefont{C.}~\bibnamefont{Cormick}},
  \bibinfo{author}{\bibfnamefont{A.}~\bibnamefont{Bermudez}},
  \bibinfo{author}{\bibfnamefont{S.~F.} \bibnamefont{Huelga}},
  \bibnamefont{and} \bibinfo{author}{\bibfnamefont{M.~B.}
  \bibnamefont{Plenio}}, \bibinfo{journal}{New Journal of Physics}
  \textbf{\bibinfo{volume}{15}}, \bibinfo{pages}{073027}
  (\bibinfo{year}{2013}).

\bibitem[{\citenamefont{Tomadin et~al.}(2011)\citenamefont{Tomadin, Diehl, and
  Zoller}}]{Tomadin11}
\bibinfo{author}{\bibfnamefont{A.}~\bibnamefont{Tomadin}},
  \bibinfo{author}{\bibfnamefont{S.}~\bibnamefont{Diehl}}, \bibnamefont{and}
  \bibinfo{author}{\bibfnamefont{P.}~\bibnamefont{Zoller}},
  \bibinfo{journal}{Phys. Rev. A} \textbf{\bibinfo{volume}{83}},
  \bibinfo{pages}{013611} (\bibinfo{year}{2011}).

\bibitem[{\citenamefont{H\"oning et~al.}(2012)\citenamefont{H\"oning, Moos, and
  Fleischhauer}}]{Honing12}
\bibinfo{author}{\bibfnamefont{M.}~\bibnamefont{H\"oning}},
  \bibinfo{author}{\bibfnamefont{M.}~\bibnamefont{Moos}}, \bibnamefont{and}
  \bibinfo{author}{\bibfnamefont{M.}~\bibnamefont{Fleischhauer}},
  \bibinfo{journal}{Phys. Rev. A} \textbf{\bibinfo{volume}{86}},
  \bibinfo{pages}{013606} (\bibinfo{year}{2012}).

\bibitem[{\citenamefont{Sieberer et~al.}(2013)\citenamefont{Sieberer, Huber,
  Altman, and Diehl}}]{Sieberer13}
\bibinfo{author}{\bibfnamefont{L.~M.} \bibnamefont{Sieberer}},
  \bibinfo{author}{\bibfnamefont{S.~D.} \bibnamefont{Huber}},
  \bibinfo{author}{\bibfnamefont{E.}~\bibnamefont{Altman}}, \bibnamefont{and}
  \bibinfo{author}{\bibfnamefont{S.}~\bibnamefont{Diehl}},
  \bibinfo{journal}{Phys. Rev. Lett.} \textbf{\bibinfo{volume}{110}},
  \bibinfo{pages}{195301} (\bibinfo{year}{2013}).

\bibitem[{\citenamefont{Carr et~al.}(2013)\citenamefont{Carr, Ritter, Adams,
  and Weatherill}}]{Carr13}
\bibinfo{author}{\bibfnamefont{C.}~\bibnamefont{Carr}},
  \bibinfo{author}{\bibfnamefont{R.}~\bibnamefont{Ritter}},
  \bibinfo{author}{\bibfnamefont{C.~S.} \bibnamefont{Adams}}, \bibnamefont{and}
  \bibinfo{author}{\bibfnamefont{K.~J.} \bibnamefont{Weatherill}},
  \bibinfo{journal}{preprint} p. \bibinfo{pages}{arXiv:1302.6621}
  (\bibinfo{year}{2013}).

\bibitem[{\citenamefont{Barreiro et~al.}(2011)\citenamefont{Barreiro, M\"uller,
  Schindler, Nigg, Monz, Chwalla, Hennrich, Roos, Zoller, and
  Blatt}}]{Barreiro11}
\bibinfo{author}{\bibfnamefont{J.}~\bibnamefont{Barreiro}},
  \bibinfo{author}{\bibfnamefont{M.}~\bibnamefont{M\"uller}},
  \bibinfo{author}{\bibfnamefont{P.}~\bibnamefont{Schindler}},
  \bibinfo{author}{\bibfnamefont{D.}~\bibnamefont{Nigg}},
  \bibinfo{author}{\bibfnamefont{T.}~\bibnamefont{Monz}},
  \bibinfo{author}{\bibfnamefont{M.}~\bibnamefont{Chwalla}},
  \bibinfo{author}{\bibfnamefont{M.}~\bibnamefont{Hennrich}},
  \bibinfo{author}{\bibfnamefont{C.}~\bibnamefont{Roos}},
  \bibinfo{author}{\bibfnamefont{P.}~\bibnamefont{Zoller}}, \bibnamefont{and}
  \bibinfo{author}{\bibfnamefont{R.}~\bibnamefont{Blatt}},
  \bibinfo{journal}{Nature} \textbf{\bibinfo{volume}{470}},
  \bibinfo{pages}{486} (\bibinfo{year}{2011}).

\bibitem[{\citenamefont{Schindler et~al.}(2013)\citenamefont{Schindler,
  M\"uller, Nigg, Barreiro, Mart\'inez, Hennrich, Monz, Diehl, Zoller, and
  Blatt}}]{Schindler13}
\bibinfo{author}{\bibfnamefont{P.}~\bibnamefont{Schindler}},
  \bibinfo{author}{\bibfnamefont{M.}~\bibnamefont{M\"uller}},
  \bibinfo{author}{\bibfnamefont{D.}~\bibnamefont{Nigg}},
  \bibinfo{author}{\bibfnamefont{J.}~\bibnamefont{Barreiro}},
  \bibinfo{author}{\bibfnamefont{E.}~\bibnamefont{Mart\'inez}},
  \bibinfo{author}{\bibfnamefont{M.}~\bibnamefont{Hennrich}},
  \bibinfo{author}{\bibfnamefont{T.}~\bibnamefont{Monz}},
  \bibinfo{author}{\bibfnamefont{S.}~\bibnamefont{Diehl}},
  \bibinfo{author}{\bibfnamefont{P.}~\bibnamefont{Zoller}}, \bibnamefont{and}
  \bibinfo{author}{\bibfnamefont{R.}~\bibnamefont{Blatt}},
  \bibinfo{journal}{Nat. Phys} \textbf{\bibinfo{volume}{9}},
  \bibinfo{pages}{361} (\bibinfo{year}{2013}).

\bibitem[{\citenamefont{Lin et~al.}(2013)\citenamefont{Lin, Gaebler, Reiter,
  Tan, Bowler, Sørensen, Leibfried, and Wineland}}]{Lin13}
\bibinfo{author}{\bibfnamefont{Y.}~\bibnamefont{Lin}},
  \bibinfo{author}{\bibfnamefont{J.~P.} \bibnamefont{Gaebler}},
  \bibinfo{author}{\bibfnamefont{F.}~\bibnamefont{Reiter}},
  \bibinfo{author}{\bibfnamefont{T.~R.} \bibnamefont{Tan}},
  \bibinfo{author}{\bibfnamefont{R.}~\bibnamefont{Bowler}},
  \bibinfo{author}{\bibfnamefont{A.~S.} \bibnamefont{Sørensen}},
  \bibinfo{author}{\bibfnamefont{D.}~\bibnamefont{Leibfried}},
  \bibnamefont{and} \bibinfo{author}{\bibfnamefont{D.~J.}
  \bibnamefont{Wineland}}, \bibinfo{journal}{preprint} p.
  \bibinfo{pages}{arXiv:1307.4443} (\bibinfo{year}{2013}).

\bibitem[{\citenamefont{Agarwal}(1970)}]{Agarwal70}
\bibinfo{author}{\bibfnamefont{G.~S.} \bibnamefont{Agarwal}},
  \bibinfo{journal}{Phys. Rev. A} \textbf{\bibinfo{volume}{2}},
  \bibinfo{pages}{2038} (\bibinfo{year}{1970}).

\bibitem[{\citenamefont{Lehmberg}(1970)}]{Lehmberg70}
\bibinfo{author}{\bibfnamefont{R.~H.} \bibnamefont{Lehmberg}},
  \bibinfo{journal}{Phys. Rev. A} \textbf{\bibinfo{volume}{2}},
  \bibinfo{pages}{883} (\bibinfo{year}{1970}).

\bibitem[{\citenamefont{Keaveney et~al.}(2012)\citenamefont{Keaveney, Sargsyan,
  Krohn, Hughes, Sarkisyan, and Adams}}]{Keaveney12}
\bibinfo{author}{\bibfnamefont{J.}~\bibnamefont{Keaveney}},
  \bibinfo{author}{\bibfnamefont{A.}~\bibnamefont{Sargsyan}},
  \bibinfo{author}{\bibfnamefont{U.}~\bibnamefont{Krohn}},
  \bibinfo{author}{\bibfnamefont{I.~G.} \bibnamefont{Hughes}},
  \bibinfo{author}{\bibfnamefont{D.}~\bibnamefont{Sarkisyan}},
  \bibnamefont{and} \bibinfo{author}{\bibfnamefont{C.~S.} \bibnamefont{Adams}},
  \bibinfo{journal}{Phys. Rev. Lett.} \textbf{\bibinfo{volume}{108}},
  \bibinfo{pages}{173601} (\bibinfo{year}{2012}).

\bibitem[{\citenamefont{Dicke}(1954)}]{Dicke54}
\bibinfo{author}{\bibfnamefont{R.~H.} \bibnamefont{Dicke}},
  \bibinfo{journal}{Phys. Rev.} \textbf{\bibinfo{volume}{93}},
  \bibinfo{pages}{99} (\bibinfo{year}{1954}).

\bibitem[{\citenamefont{Malossi et~al.}(2013)\citenamefont{Malossi, Valado,
  Scotto, Huillery, Pillet, Ciampini, Arimondo, and Morsch}}]{Malossi13}
\bibinfo{author}{\bibfnamefont{N.}~\bibnamefont{Malossi}},
  \bibinfo{author}{\bibfnamefont{M.}~\bibnamefont{Valado}},
  \bibinfo{author}{\bibfnamefont{S.}~\bibnamefont{Scotto}},
  \bibinfo{author}{\bibfnamefont{P.}~\bibnamefont{Huillery}},
  \bibinfo{author}{\bibfnamefont{P.}~\bibnamefont{Pillet}},
  \bibinfo{author}{\bibfnamefont{D.}~\bibnamefont{Ciampini}},
  \bibinfo{author}{\bibfnamefont{E.}~\bibnamefont{Arimondo}}, \bibnamefont{and}
  \bibinfo{author}{\bibfnamefont{O.}~\bibnamefont{Morsch}},
  \bibinfo{journal}{preprint} p. \bibinfo{pages}{arXiv:1308.1854}
  (\bibinfo{year}{2013}).

\bibitem[{\citenamefont{Lee et~al.}(2011)\citenamefont{Lee, H\"affner, and
  Cross}}]{Lee11}
\bibinfo{author}{\bibfnamefont{T.~E.} \bibnamefont{Lee}},
  \bibinfo{author}{\bibfnamefont{H.}~\bibnamefont{H\"affner}},
  \bibnamefont{and} \bibinfo{author}{\bibfnamefont{M.~C.} \bibnamefont{Cross}},
  \bibinfo{journal}{Phys. Rev. A} \textbf{\bibinfo{volume}{84}},
  \bibinfo{pages}{031402} (\bibinfo{year}{2011}).

\bibitem[{\citenamefont{Lee et~al.}(2012)\citenamefont{Lee, Haffner, and
  Cross}}]{Lee12}
\bibinfo{author}{\bibfnamefont{T.~E.} \bibnamefont{Lee}},
  \bibinfo{author}{\bibfnamefont{H.}~\bibnamefont{Haffner}}, \bibnamefont{and}
  \bibinfo{author}{\bibfnamefont{M.~C.} \bibnamefont{Cross}},
  \bibinfo{journal}{Phys. Rev. Lett.} \textbf{\bibinfo{volume}{108}},
  \bibinfo{pages}{023602} (\bibinfo{year}{2012}).

\bibitem[{\citenamefont{Ates et~al.}(2012)\citenamefont{Ates, Olmos, Garrahan,
  and Lesanovsky}}]{Ates12-2}
\bibinfo{author}{\bibfnamefont{C.}~\bibnamefont{Ates}},
  \bibinfo{author}{\bibfnamefont{B.}~\bibnamefont{Olmos}},
  \bibinfo{author}{\bibfnamefont{J.~P.} \bibnamefont{Garrahan}},
  \bibnamefont{and}
  \bibinfo{author}{\bibfnamefont{I.}~\bibnamefont{Lesanovsky}},
  \bibinfo{journal}{Phys. Rev. A} \textbf{\bibinfo{volume}{85}},
  \bibinfo{pages}{043620} (\bibinfo{year}{2012}).

\bibitem[{\citenamefont{Lesanovsky et~al.}(2013)\citenamefont{Lesanovsky, van
  Horssen, Gu\ifmmode \mbox{\c{t}}\else \c{t}\fi{}\ifmmode~\u{a}\else
  \u{a}\fi{}, and Garrahan}}]{Lesanovsky12b}
\bibinfo{author}{\bibfnamefont{I.}~\bibnamefont{Lesanovsky}},
  \bibinfo{author}{\bibfnamefont{M.}~\bibnamefont{van Horssen}},
  \bibinfo{author}{\bibfnamefont{M.~u. u. u.~u.} \bibnamefont{Gu\ifmmode
  \mbox{\c{t}}\else \c{t}\fi{}\ifmmode~\u{a}\else \u{a}\fi{}}},
  \bibnamefont{and} \bibinfo{author}{\bibfnamefont{J.~P.}
  \bibnamefont{Garrahan}}, \bibinfo{journal}{Phys. Rev. Lett.}
  \textbf{\bibinfo{volume}{110}}, \bibinfo{pages}{150401}
  (\bibinfo{year}{2013}).

\bibitem[{\citenamefont{Olmos et~al.}(2013)\citenamefont{Olmos, Yu, Singh,
  Schreck, Bongs, and Lesanovsky}}]{Olmos13}
\bibinfo{author}{\bibfnamefont{B.}~\bibnamefont{Olmos}},
  \bibinfo{author}{\bibfnamefont{D.}~\bibnamefont{Yu}},
  \bibinfo{author}{\bibfnamefont{Y.}~\bibnamefont{Singh}},
  \bibinfo{author}{\bibfnamefont{F.}~\bibnamefont{Schreck}},
  \bibinfo{author}{\bibfnamefont{K.}~\bibnamefont{Bongs}}, \bibnamefont{and}
  \bibinfo{author}{\bibfnamefont{I.}~\bibnamefont{Lesanovsky}},
  \bibinfo{journal}{Phys. Rev. Lett.} \textbf{\bibinfo{volume}{110}},
  \bibinfo{pages}{143602} (\bibinfo{year}{2013}).

\bibitem[{\citenamefont{M\o{}lmer et~al.}(1993)\citenamefont{M\o{}lmer, Castin,
  and Dalibard}}]{Molmer93}
\bibinfo{author}{\bibfnamefont{K.}~\bibnamefont{M\o{}lmer}},
  \bibinfo{author}{\bibfnamefont{Y.}~\bibnamefont{Castin}}, \bibnamefont{and}
  \bibinfo{author}{\bibfnamefont{J.}~\bibnamefont{Dalibard}},
  \bibinfo{journal}{J. Opt. Soc. Am. B} \textbf{\bibinfo{volume}{10}},
  \bibinfo{pages}{524} (\bibinfo{year}{1993}).

\bibitem[{\citenamefont{Dalibard et~al.}(1992)\citenamefont{Dalibard, Castin,
  and M\o{}lmer}}]{Dalibard92}
\bibinfo{author}{\bibfnamefont{J.}~\bibnamefont{Dalibard}},
  \bibinfo{author}{\bibfnamefont{Y.}~\bibnamefont{Castin}}, \bibnamefont{and}
  \bibinfo{author}{\bibfnamefont{K.}~\bibnamefont{M\o{}lmer}},
  \bibinfo{journal}{Phys. Rev. Lett.} \textbf{\bibinfo{volume}{68}},
  \bibinfo{pages}{580} (\bibinfo{year}{1992}).

\bibitem[{\citenamefont{Zhou et~al.}(2010)\citenamefont{Zhou, Xu, Chen, and
  Chen}}]{Zhou10}
\bibinfo{author}{\bibfnamefont{X.}~\bibnamefont{Zhou}},
  \bibinfo{author}{\bibfnamefont{X.}~\bibnamefont{Xu}},
  \bibinfo{author}{\bibfnamefont{X.}~\bibnamefont{Chen}}, \bibnamefont{and}
  \bibinfo{author}{\bibfnamefont{J.}~\bibnamefont{Chen}},
  \bibinfo{journal}{Phys. Rev. A} \textbf{\bibinfo{volume}{81}},
  \bibinfo{pages}{012115} (\bibinfo{year}{2010}).

\bibitem[{\citenamefont{Krischek et~al.}(2011)\citenamefont{Krischek,
  Schwemmer, Wieczorek, Weinfurter, Hyllus, Pezz\'e, and Smerzi}}]{Krischek11}
\bibinfo{author}{\bibfnamefont{R.}~\bibnamefont{Krischek}},
  \bibinfo{author}{\bibfnamefont{C.}~\bibnamefont{Schwemmer}},
  \bibinfo{author}{\bibfnamefont{W.}~\bibnamefont{Wieczorek}},
  \bibinfo{author}{\bibfnamefont{H.}~\bibnamefont{Weinfurter}},
  \bibinfo{author}{\bibfnamefont{P.}~\bibnamefont{Hyllus}},
  \bibinfo{author}{\bibfnamefont{L.}~\bibnamefont{Pezz\'e}}, \bibnamefont{and}
  \bibinfo{author}{\bibfnamefont{A.}~\bibnamefont{Smerzi}},
  \bibinfo{journal}{Phys. Rev. Lett.} \textbf{\bibinfo{volume}{107}},
  \bibinfo{pages}{080504} (\bibinfo{year}{2011}).

\bibitem[{\citenamefont{Ostermann et~al.}(2013)\citenamefont{Ostermann, Ritsch,
  and Genes}}]{Ostermann13}
\bibinfo{author}{\bibfnamefont{L.}~\bibnamefont{Ostermann}},
  \bibinfo{author}{\bibfnamefont{H.}~\bibnamefont{Ritsch}}, \bibnamefont{and}
  \bibinfo{author}{\bibfnamefont{C.}~\bibnamefont{Genes}},
  \bibinfo{journal}{preprint} p. \bibinfo{pages}{arXiv:1307.2558}
  (\bibinfo{year}{2013}).

\end{thebibliography}
\end{document}